\documentclass{elsart}
\usepackage{amsmath}
\usepackage{amssymb}
\begin{document}
\begin{frontmatter}
\title{A deterministic Bell model}
\author{Samuel Colin}
\address{University of Louvain-la-Neuve, FYMA,
Chemin du cyclotron 2, \mbox{B-1348 Louvain-la-Neuve}, Belgium}
\ead{colin@fyma.ucl.ac.be}
\begin{abstract}
Following ideas given by John Bell in a paper entitled \textit{Beables for quantum field theory}, we show that it is
possible to obtain a realistic and deterministic interpretation of any quantum field-theoretic model involving Fermi fields.
\end{abstract}
\end{frontmatter}
\section{Introduction}
At the heart of the orthodox interpretation of the quantum theory is the measurement postulate. We recall the context in which this
postulate has been introduced and why it is unsatisfactory\footnote{For more complete reviews of the
measurement problem, see \cite{bell2} and \cite{bricmont1}.}. It supposes that the wave-function is the complete description of a
system of particles. If we assume that the wave-function evolves according to a linear
equation, we face the problem of macroscopic superpositions (Schr\"odinger's cat)\footnote{At this point, it should be
stressed that superpositions have an absolute meaning, since any measurement is finally a measurement of positions.}. Since those
macroscopic superpositions are not observed, the measurement postulate is introduced, in order to stop the branching process
somewhere, where systems encounter observers, who force them to collapse. By introducing the collapse, the world is split into two
parts: a quantum world, made of systems, which can be in superpositions and are unable to perform measurements, and a
classical world, made of observers, who on the contrary have the ability to cause a collapse and do not enter in superpositions.
Systems are described by wave functions, whereas observers are described by positions. If one asks where the boundary between
these two worlds is to be found or even why there is a boundary, the orthodox interpretation gives no precise answer. Nevertheless,
such ill-defined concepts appear in the fundamental postulates of the orthodox interpretation. The second argument against the
measurement is that the collapse postulate clashes with the reductionist project that has always guided the physicists: it has
simply no meaning to speak of a wave-function of the universe in the Copenhagen interpretation. Would there be a theory that
explains how the collapse appears for macroscopic objects, there would be nothing to worry about. There are interpretations
following that way, but they imply modifications of the Schr\"odinger equation, for example the Ghirardi-Rimini-Weber
theory. On the contrary of the orthodox interpretation, it has no need for a vague division of the world into quantum and classical
parts. Another attitude towards superpositions is that of Everett (see \cite{bell2}, chapter 11,15  for a critical review).

Hidden-variables theories constitute the second category of interpretations: these are interpretations in which the
wave-function is not the complete description with a system of particles. Now, the widespread claim is that those interpretations
are ruled out by Bell's inequality and the experiments that have been carried out later. In fact local ones are ruled out, but
since non-locality is commonly claimed to be unacceptable (even weak non-locality, which does not lead to paradoxical situations),
it is said that any hidden-variables theory is incompatible with the quantum theory\footnote{A theory is said to be non-local if it
predicts that a cause and its effect are simultaneous in an inertial frame (even if it is impossible to distinguish the cause
from the effect).}. That is the wrong way to present the theoretical situation. To present it correctly, we have to return
to the EPR paradox, which in essence says that some quantum correlations cannot be explained in a local way, unless we say
that the quantum theory is incomplete. Since local hidden-variables theories are ruled out by Bell's inequality and
experiments, the only way to explain quantum correlations is to revert to non-locality (which is in fact hidden in the collapse
postulate). Then, to suppress the ill-defined measurement postulate, it is preferable to interpret the quantum world by a
non-local hidden-variables theory. Such an interpretation exists for non-relativistic quantum mechanics: it is the de Broglie-Bohm
pilot wave theory, whose John Bell, who is often credited of the refutation of hidden variables theories, has been one of the main
advocates.

Pilot-wave theory is a realistic theory, in so far as the positions of the particles exist and are simply revealed by
position measurements (the positions are beables, a term coined by Bell). If there are $n$ particles, the universe is thus completely
described by the couple $(\vec{X}(t),\Psi(t,\vec{X}))$, where $\vec{X}$ is a point in a configuration space of dimension $3n$.
$\Psi(t,\vec{X})$ evolves according to the Schr\"odinger equation, whereas the equation of motion for $\vec{X}(t)$ is such that if we
consider a set of universes with the same wave-function and initial configurations chosen according to the probability density
$|\Psi(t_0,\vec{X})|^2$, then the final configurations will be distributed according to $|\Psi(t,\vec{X})|^2$ for any later time
$t$ (see \cite{goldstein2} for a survey of the non-relativistic pilot wave theory or \cite{holland} for a complete study).

The question that comes to mind is to know if the quantum field theory can also be interpreted as a non-local hidden-variable
theory. At the time Bell wrote his paper \cite{bell1}, Bohm had already shown that it was possible to build a realistic
interpretation of any bosonic quantum field theory \cite{bohm1}. To achieve that goal, Bohm took the field as the beable, however
he was not able to do the same for fermions. The aim of Bell was then to show that it was also possible to build a realistic
interpretation of any fermionic quantum field theory, along the pilot-wave ideas. Bell managed doing so but he took a different
beable: the fermion number density. It is slightly different from the non-relativistic pilot-wave theory, whose beables are the
positions of the particles. The model is also formulated on a spatial lattice (space is discrete but time remains continuous).
His model is stochastic, but he suspected that the theory would become deterministic in the continuum limit.

This paper is organized in the following way. First, we will give a brief presentation of the lattice Bell model. Then we will ask
ourselves what is the physical content of the Bell model, by studying the fermion-number density. Finally, we will show that
the stochastic Bell model can be turned into a deterministic one.
\section{The Bell model}
The Bell model is defined on a finite lattice, whose sites are labelled by an index
\begin{equation*}
l=1,2,\ldots,L~,
\end{equation*}
where $L$ is very large. The lattice fermion-number operator is
defined by
\begin{equation}\label{fnd}
\psi^\dagger(l)\psi(l)=\sum_{a=1}^{a=4}\psi^\dagger_a(l)\psi_a(l)~,
\end{equation}
where $\psi(l)$ is a four-component lattice Dirac field (we only
consider a single species of fermions here). Since
$[\psi^\dagger(l)\psi(l),\psi^\dagger(k)\psi(k)]=0$, for $k\neq
l$, it is possible to define eigenstates of the fermion-number
density; those eigenstates are defined by $|nq\rangle$
\begin{equation*}
\psi^\dagger(l)\psi(l)|nq\rangle=F(l)|nq\rangle~,
\end{equation*}
where $q$ are eigenvalues of observables $Q$ such that
\begin{equation*}
\{\psi^\dagger(1)\psi(1),\ldots,\psi^\dagger(L)\psi(L),Q\}
\end{equation*}
is a complete set of commuting observables, and $n$ is a
fermion-number density configuration ($n=(F(1),F(2)\ldots,F(L))$).
We will see that the $F(l)$ belong to $\{0,1,2,3,4\}$. The
fermion-number operator is
$F=\displaystyle\sum_{l}\psi^\dagger(l)\psi(l)$.

Since it is impossible to build a no hidden-variables theorem
concerning the fermion-number density (see \cite{bricmont1} for a
hint of the proof), the fermion-number density can be given the
beable status. That means that any measurement of the
fermion-number density at time $t$ would simply reveal a
pre-existing value
\begin{equation*}
n(t)=\{n_1(t),n_2(t),\ldots,n_L(t)\}~.
\end{equation*}
This is the first element of reality in the description of the
universe; the second element is the pilot-state $|\Psi(t)\rangle$,
which always evolves according to the Schr\"odinger equation
\begin{equation*}
i\frac{\partial |\Psi(t)\rangle}{\partial t}=H|\Psi(t)\rangle~,
\end{equation*}
where $H$ is the hamiltonian. Thus the complete description of the
universe at time $t$ would be given by the couple
$(n(t),|\Psi(t)\rangle)$.

To complete the model, one must say how the real fermion-number
density $n(t)$ evolves in time. The equation of motion for $n(t)$,
called the velocity-law, must be such that the predictions of the
orthodox quantum field theory are regained. Let us define $P_n(t)$
as the probability for the universe to BE in configuration $n$ at
time $t$. Since a measurement of the fermion-number density at
time $t$ reveals the pre-existing value $n(t)$, the following
relation
\begin{equation}\label{equi}
P_n(t)=\sum_{q}|\langle n,q|\Psi(t)\rangle|^2
\end{equation}
must be satisfied. Bell gives a stochastic equation of motion that
reproduces eq. (\ref{equi}), provided that the initial
fermion-number density configurations (at time $t_0$) are
distributed according to
\begin{equation*}
P_n(t_0)=\sum_{q}|\langle n,q|\Psi(t_0)\rangle|^2~.
\end{equation*}
This stochastic velocity-law is defined as follows. Let us take
$T_{nm}(t)$ to be the jump-rate, at time $t$, to configuration
$n$, if the universe is in configuration $m$ at time $t$. If one
takes
\begin{align}\label{svl}
T_{nm}=&J_{nm}/D_m~,&\\
J_{nm}=&2\sum_{q\;\!p}\mathfrak{Re}[\langle\Psi(t)|nq\rangle\langle
nq|-iH|mp\rangle\langle mp|\Psi(t)\rangle]~,&\nonumber\\
D_{m}=&\sum_{q}|\langle mq|\Psi(t)\rangle|^2~,&\nonumber
\end{align}
if $J_{nm}>0$, and $T_{nm}=0$ otherwise, then it can be shown that
the predictions of the orthodox quantum field theory are
regained\footnote{The idea is to take the time-derivative of eq.
(\ref{equi}) and to show that it is indeed verified (see
\cite{bell1}).}. This constitutes the Bell model.
\section{\label{sec:fdvp}Further developments of the Bell model}
In the continuum limit (lattice-spacing going to zero), the
fermion-number density (eq. (\ref{fnd})) would become
$\psi^\dagger(\vec{x})\psi(\vec{x})$, so the fermion-number
density is directly related to the charge-density operator, except
that no normal-ordering is implied in the definition of the
fermion-number density (otherwise its eigenvalues would not be
positive). It is worth mentioning that it is impossible to build
eigenstates of the fermion-number density which are also
eigenstates of the particle-number (number of electrons plus
number of positrons). The proof (that it is possible to find test
functions $f$ such that
\begin{equation*}
[\int d^3\vec{x} f(\vec{x})
\psi^\dagger(\vec{x})\psi(\vec{x}),N]\neq 0~,
\end{equation*}
where $N$ is the particle number) is given in the appendix A of
\cite{colin}. Then one may ask why this operator is called the
fermion-number density; the answer will be given in the next
section.

Let us return to the lattice case. We assume that the lattice
fermionic fields satisfy the following canonical anti-commutation
relations
\begin{align}\label{car}
&\{\psi_a(k),\psi_b(l)\}=0&
&\{\psi_a(k),\psi^\dagger_b(l)\}=\delta_{ab}\delta_{lk}~,&
\end{align}
with $l,k\in\{1,2,\ldots,L\}$, $a,b\in\{1,2,3,4\}$, and we want to
construct the eigenstates of the fermion-number density. For that
purpose, we start from a state $|\Phi\rangle$ such that
\begin{equation*}
\psi^\dagger(l)\psi(l)|\Phi\rangle=F(l)|\Phi\rangle
\end{equation*}
and we apply an operator $\psi^\dagger_b(k)$ on it. From the
relations (\ref{car}), it can be seen that
\begin{equation*}
\sum_{a}\psi^\dagger_a(l)\psi_a(l)\psi^\dagger_b(k)|\Phi\rangle=(\delta_{kl}+F(l))\psi^\dagger_b(k)|\Phi\rangle~,
\end{equation*}
so that any of the four operators $\psi^\dagger_b(k)$ ($k$ fixed)
creates a quantum of the fermion-number at site $k$. In the same
way, the operators $\psi_b(k)$ are annihilators of the
fermion-number at site $k$. To obtain the eigenstates of the
fermion-number density, we thus have to start from a state
$|v\rangle$ such that
\begin{equation*}
\psi_a(l)|v\rangle=0\;\;\forall\,a
\in\{1,2,3,4\}\;\;\forall\,l\in\{1,2,\ldots,L\}
\end{equation*}
and apply various creators on it. Since $\psi_a(l)$ contains
operators that destroy electrons and operators that create
positrons, the state $|v\rangle$ is simply the positronic sea
$|ps\rangle$ (all the positron states occupied):
\begin{equation*}
\psi_a(l)|v\rangle=\psi_a(l)|ps\rangle=
\psi_a(l)\prod_{\vec{p},s}d^\dagger_s(\vec{p})|0_1\rangle=0
\;\;\forall\,a,l~,
\end{equation*}
where $|0_1\rangle$ is the usual vacuum (no electron, no
positron), $s$ are helicity symbols and $d^\dagger_s(\vec{p})$ is
the operator that creates a positron of momentum $\vec{p}$, energy
$\sqrt{|\vec{p}|^2+m^2}$ and helicity $(-1)^{s+1}$.

Let us consider an example: an eigenstate of the fermion-number
density with fermion-number equal to 2, with one quantum localized
at site $l_1$ and another at site $l_2$. There are 16 eigenstates
of the fermion-number density with that eigenvalue
($F(l)=\delta_{ll_1}+\delta_{ll_2}$), namely
\begin{align*}
&\psi^\dagger_{a_1}(l_1)\psi^\dagger_{a_2}(l_2)|ps\rangle& &a_1,a_2\in\{1,2,3,4\}~.&
\end{align*}
It should be stressed again that these eigenstates are not
eigenstates of the particle number and that it is generally
impossible to take particular combinations of them in order to
obtain eigenstates of the particle number, the two exceptions
being the positronic sea and the electronic sea.

Since $[H,F]=0$, the fermion-number is conserved and in accordance
with the well-known super-selection rule that forbids
superpositions of states with different eigenvalues of the
fermion-number, we have that
\begin{equation*}
F|\Psi(t)\rangle=\omega|\Psi(t)\rangle~,
\end{equation*}
where $\omega$ is a positive integer belonging to
$\{0,1,\ldots,4L\}$. Thus $|\Psi(t)\rangle$ can always be
decomposed along the eigenstates of the fermion-number density
with fermion-number equal to $\omega$. Those eigenstates are
\begin{align*}
&\psi^\dagger_{a_1}(l_1)\ldots\psi^\dagger_{a_\omega}(l_\omega)|ps\rangle&
&a_1,\ldots,a_\omega\in\{1,2,3,4\},\,l_1,\ldots,l_\omega\in\{1,2,\ldots,L\}~.&
\end{align*}
Then
\begin{equation*}
|\Psi(t)\rangle=\frac{1}{\omega!}\sum_{a_1\,l_1}\ldots\sum_{a_\omega\,l_\omega}
\Psi_{a_1\ldots a_\omega}(t,l_1,\ldots,l_\omega)
\psi^\dagger_{a_1}(l_1)\ldots\psi^\dagger_{a_\omega}(l_\omega)|ps\rangle~,
\end{equation*}
where the lattice wave-function $\Psi_{a_1\ldots
a_\omega}(t,l_1,\ldots,l_\omega)$ is antisymmetric under any
transposition of two labels belonging to $\{1,\ldots,\omega\}$.

Physically, $\omega$ is the number of negative charges that one
has to add to the charge contained in the positronic sea ($2eL$)
in order to obtain the charge contained in $|\Psi(t)\rangle$. In
the Bell ontology, these $\omega$ negative charges jump from site
to site according to the stochastic velocity-law (eq.~(\ref{svl}))
and a measurement of the fermion-number density simply reveals
their positions (or more exactly the sites that they occupy). If
the fermion-number density is sufficient to describe the outputs
of the measuring devices, then we have a realistic interpretation
of any lattice quantum field theory involving Fermi fields. Due to
the simple relation between fermion-number density and charge
density, it is also strictly equivalent to say that the charge
density is the beable.

Now we ask the question whether it is possible that the Bell model
becomes deterministic in the continuum limit (lattice spacing
going to zero) or whether it is possible to turn the stochastic
Bell model into a deterministic one. There are two strong
arguments in favor of that conjecture; the deterministic character
of the Schr\"odinger equation and the conservation of the
fermion-number.
\section{The continuum Dirac quantum field theory}
In the continuum limit, the finite lattice is replaced by a cubic
box of volume $V$ and the fermion-number density operator becomes
\begin{equation*}
\psi^\dagger(\vec{x})\psi(\vec{x})=\sum_{a=1}^{a=4}\psi^\dagger_a(\vec{x})\psi_a(\vec{x})~,
\end{equation*}
The components of the Dirac fields satisfy the following
canonical anti-commutation relations
\begin{align}\label{car2}
&\{\psi^\dagger_a(\vec{x}),\psi_b(\vec{y})\}=\delta_{ab}\delta^3(\vec{x}-\vec{y})&
&\{\psi_a(\vec{x}),\psi_b(\vec{y})\}=0~.&
\end{align}
Integrating the fermion-number density over the cubic box of
volume $V$, we get the fermion-number
\begin{equation}\label{fn}
F=\int_{V}d^3\vec{x} \psi^\dagger(\vec{x})\psi(\vec{x})~.
\end{equation}
The free hamiltonian is defined by
\begin{equation}\label{fh}
H_0=\int_{V}d^3\vec{x}
\psi^\dagger(\vec{x})[-i\vec{\alpha}\cdot\vec{\nabla}+m\beta]\psi(\vec{x})~.
\end{equation}
Let us take
$u_s(\vec{p})e^{-iE_{\vec{p}}t}e^{i\vec{p}\cdot\vec{x}}/\sqrt{V}$
to be a free solution of the classical Dirac equation with
momentum $\vec{p}$, energy $E_{\vec{p}}=\sqrt{|\vec{p}|^2+m^2}$
and helicity $(-1)^{s+1}$, and
$v_s(\vec{p})e^{iE_{\vec{p}}t}e^{-i\vec{p}\cdot\vec{x}}/\sqrt{V}$
to be a free solution of the classical Dirac equation with
momentum $-\vec{p}$, energy $-E_{\vec{p}}$ and helicity $(-1)^s$.
For the free solutions to be orthogonal and normalized in a
covariant way, the following relations
\begin{align}\label{orth}
&u^\dagger_s(\vec{p})u_r(\vec{p})=\delta_{rs}\frac{E_{\vec{p}}}{m}&
&u^\dagger_s(\vec{p})v_r(-\vec{p})=0&
&v^\dagger_s(\vec{p})v_r(\vec{p})=\delta_{rs}\frac{E_{\vec{p}}}{m}&
\end{align}
must be satisfied. Since the quantum field $\psi$ is a solution of
the Dirac equation, it is a superposition of free solutions with
operators as coefficients; taking
\begin{equation}\label{sf}
\psi(t,\vec{x})=\sqrt{\frac{1}{V}}\sum_{s}\sum_{\vec{p}}\sqrt{\frac{m}{E_{\vec{p}}}}
[c_s(\vec{p})u_s(\vec{p})e^{-iE_{\vec{p}}t}e^{i\vec{p}\cdot\vec{x}}+
d^\dagger_s(\vec{p})v_s(\vec{p})e^{iE_{\vec{p}}t}e^{-i\vec{p}\cdot\vec{x}}]~,
\end{equation}
with
$\{c_s(\vec{p}),c^\dagger_r(\vec{q})\}=\delta^r_s\delta_{\vec{p}\;\!\vec{q}}$,
$\{d_s(\vec{p}),d^\dagger_r(\vec{q})\}=\delta^r_s\delta_{\vec{p}\;\!\vec{q}}$
and all other anti-com\-mutators vanishing, the anti-commutation
relations (\ref{car2}) are regained.

The vacuum is still called $|0_1\rangle$. From the expressions of
the various observables in the momentum space, it appears that the
operator $c^\dagger_s(\vec{p})$ ($d^\dagger_s(\vec{p})$) creates
an electron (a positron) of momentum $\vec{p}$, energy
$E_{\vec{p}}$ and helicity $(-1)^{s+1}$.

With the help of eq. (\ref{orth}) and eq. (\ref{sf}), we find that
the expression of the fermion-number (eq. (\ref{fn})) in the
momentum space is
\begin{equation*}
F=\sum_{s=1}^{s=2}\sum_{\vec{p}}[c^\dagger_s(\vec{p})c_s(\vec{p})+d_s(\vec{p})d^\dagger_s(\vec{p})]~.
\end{equation*}
Hence, there is only one eigenstate of the fermion-number with the
lowest eigenvalue (zero), it is the positronic sea $|ps\rangle$.
The eigenstates of the fermion-number density are obtained from
the positronic sea by applying creators $\psi^\dagger_a(\vec{x})$
on it. For example, there are $16$ eigenstates of the
fermion-number density with eigenvalue
$f(\vec{x})=\delta^3(\vec{x}-\vec{x}_1)+\delta^3(\vec{x}-\vec{x}_2)$;
those eigenstates are
\begin{equation*}
\psi^\dagger_{a_1}(\vec{x}_1)\psi^\dagger_{a_2}(\vec{x}_2)|ps\rangle~~a_1,a_2\in\{1,2,3,4\}~.
\end{equation*}
Instead of working with positrons, we can keep the states of
negative energy. This can be accomplished by making the
substitutions
\begin{align*}
&d^\dagger_s(\vec{p})\rightarrow \zeta_s(-\vec{p})&
&d_s(\vec{p})\rightarrow \zeta^\dagger_s(-\vec{p})~,&
\end{align*}
where $\zeta^\dagger_s(\vec{p})$ is the operator that creates an
electron of momentum $\vec{p}$, energy $-E_{\vec{p}}$ and helicity
$(-1)^{s+1}$. This is Dirac's prescription (a hole in the Dirac
sea is equivalent to a positron). Then, if that interpretation is
used, another vacuum has to be defined; we call it $|0_2\rangle$
\begin{align*}
&c_s(\vec{p})|0_2\rangle=0& &\zeta_s(\vec{p})|0_2\rangle=0&
&\forall~s,\vec{p}~.&
\end{align*}
This vacuum $|0_2\rangle$ is not the usual vacuum $|0_1\rangle$;
in fact, $|0_1\rangle$ is equivalent to the Dirac sea, and can be
obtained from $|0_2\rangle$ by filling all the negative-energy
states. The expression of the fermion-number in the momentum space
becomes
\begin{equation*}
F=\sum_{s=1}^{s=2}\sum_{\vec{p}}[c^\dagger_s(\vec{p})c_s(\vec{p})+\zeta^\dagger_s(\vec{p})\zeta_s(\vec{p})]~,
\end{equation*}
so that the fermion-number is the number of electrons, but of
positive and negative energy. Following that interpretation, the
fermion-number is really what its name implies and the beables of
the Bell model would be the positions of the electrons (but of
positive and negative energy). The state with lowest
fermion-number, destroyed by any annihilator $\psi_a(\vec{x})$, is
the vacuum $|0_2\rangle$. The signification of the operator
$\psi^\dagger_a(\vec{x})$ is that it creates an electron localized
at point $\vec{x}$. But the state
$\zeta^\dagger_s(\vec{p})|0_2\rangle$ has no direct
interpretation, showing that the negative-energy electrons are not
appropriated to the study of properties related to the momentum
space. The point we want to make is that these negative-energy
states are well suited to the study of localized properties. Let
us show it. We start from the Schr\"odinger equation
\begin{equation*}
i\hbar\frac{\partial|\Psi(t)\rangle}{\partial
t}=H_0|\Psi(t)\rangle~.
\end{equation*}
Since $[H_0,F]=0$, and in accordance with the super-selection
rule, we know that $|\Psi(t)\rangle$ is an eigenstate of the
fermion-number. Let us consider the case where there is only one
quantum of the fermion-number:
\begin{equation*}
F|\Psi(t)\rangle=|\Psi(t)\rangle~.
\end{equation*}
Then $|\Psi(t)\rangle$ can be decomposed along the eigenstates of
the fermion-number density with fermion-number equal to $1$, which
are
\begin{equation*}
\psi^\dagger_a(\vec{x})|0_2\rangle\quad (\vec{x}\in\mathbb{R}^3,~a\in\{1,2,3,4\})~,
\end{equation*}
since $|ps\rangle=|0_2\rangle$. Thus, in our case,
\begin{equation*}
|\Psi(t)\rangle=\sum_{a}\int
d^3\vec{x}\Psi_a(t,\vec{x})\psi^\dagger_a(\vec{x})|0_2\rangle~.
\end{equation*}
Inserting the previous equation in the Schr\"odinger equation,
using the eq. (\ref{car2}), and the definition of the hamiltonian
(eq. (\ref{fh})), one finds that
\begin{equation*}
i\frac{\partial\Psi(t,\vec{x})}{\partial
t}=-i\vec{\alpha}\cdot\vec{\nabla}\Psi(t,\vec{x})+m\beta\Psi(t,\vec{x})~,
\end{equation*}
which is the Dirac equation. So the link is made between the first
and the second quantization. And that shows that the beables of
the Bell model are the same hidden variables that are used in the
Bohm theory for Dirac particles (first quantization). This is
self-consistent since the same observable (fermion-number density)
is used in both theories.

All the occurrences of the positronic sea $|ps\rangle$ will now be
replaced by the vacuum $|0_2\rangle$, since this point of view
seems more fundamental.

From now on, the hamiltonian is not restricted to the free case
($H_0\rightarrow H=H_0+H_{I}$ and the interaction terms are made
of the Dirac field, for instance $H_{I}=g(\bar{\psi}\psi)^2$,
where $g$ is a coupling constant). Since $[H,F]=0$, the
pilot-state is an eigenstate of the fermion-number
(super-selection rule) and the corresponding eigenvalue is still
denoted by $\omega$:
\begin{equation*}
\int
d^3\vec{x}\psi^\dagger(\vec{x})\psi(\vec{x})|\Psi(t)\rangle=\omega|\Psi(t)\rangle~.
\end{equation*}
A problem is that in the continuum limit, $\omega$ becomes
infinite for any state containing a finite number of electrons and
positrons. But let us consider the case $\omega$ finite for the
moment (it will be of interest for us in the next section).

Thus $|\Psi(t)\rangle$ can be decomposed along the eigenstates of
the fermion-number density with fermion-number equal to $\omega$:
\begin{equation*}
|\Psi(t)\rangle=\frac{1}{\omega!}\sum_{a_1=1}^{a_1=4}\cdot\cdot\sum_{a_\omega=1}^{a_\omega=4}\int
d^3\vec{x}_1\cdot\cdot d^3\vec{x}_\omega\Psi_{a_1\cdot\cdot
a_\omega}(t,\vec{x}_1,\cdot\cdot,\vec{x}_\omega)
\psi^\dagger_{a_1}(\vec{x}_1)\cdot\cdot\psi^\dagger_{a_\omega}(\vec{x}_\omega)
|0_2\rangle~,
\end{equation*}
where the wave function $\Psi_{a_1\cdot\cdot
a_\omega}(t,\vec{x}_1,\cdot\cdot,\vec{x}_\omega)$ is
antisymmetric, under any transposition of two of the labels $1$ to
$\omega$.

In the standard interpretation, the probability density to observe
the universe in a configuration
$(\vec{x}_1,\cdot\cdot,\vec{x}_\omega)$ is
\begin{equation}\label{pd}
\rho(t,\vec{x}_1,\cdot\cdot,\vec{x}_\omega)=
\sum_{a_1=1}^{a_1=4}\cdot\cdot\sum_{a_\omega=1}^{a_\omega=4}
|\langle\Psi(t)|\psi^\dagger_{a_1}(\vec{x_1})\cdot\cdot\psi^\dagger_{a_\omega}
(\vec{x_\omega})|0_2\rangle|^2~,
\end{equation}
and we have the relation
\begin{equation}\label{intrho}
\int d^3{\vec{x}_1}\cdot\cdot
d^3{\vec{x}_\omega}\rho(t,\vec{x}_1,\cdot\cdot,\vec{x}_\omega)=1~.
\end{equation}
Its time-derivative could be deduced from the relation
\begin{equation}\label{co}
\frac{\partial\rho(t,\vec{X})}{\partial
t}+\vec{\nabla}\cdot\vec{J}(t,\vec{X})=0~,
\end{equation}
where $\vec{X}=(\vec{x}_1,\cdot\cdot,\vec{x}_\omega)$ and
$\vec{J}$ is a probability density current in the configuration
space of dimension $3\omega$, provided that the fields go to zero
fast enough as $|\vec{X}|\rightarrow 0$.
Let us consider the free case for the moment ($H$ is the free
Dirac hamiltonian $H_0$). The time derivative of eq.
(\ref{intrho}) gives the relation:
\begin{align*}
&\frac{d}{dt}\int d^3{\vec{x}_1}\ldots
d^3{\vec{x}_\omega}\rho(t,\vec{x}_1,\ldots,\vec{x}_\omega)=
\sum_{a_1=1}^{a_1=4}\ldots\sum_{a_\omega=1}^{a_\omega=4} \int
d^3\vec{x}_1\ldots d^3\vec{x}_\omega&\\&\frac{d}{dt}[
\langle\Psi(t_0)|\psi^\dagger_{a_1}(\vec{x_1},t)\ldots\psi^\dagger_{a_\omega}(\vec{x}_\omega,t)|0_2\rangle
\langle
0_2|\psi_{a_\omega}(\vec{x}_\omega,t)\ldots\psi_{a_1}(\vec{x_1},t)|\Psi(t_0)\rangle]=0~,&
\end{align*}
where we have switched to the Heisenberg picture. It can be
simplified, knowing that
\begin{equation*}
i\frac{\partial\psi(t,\vec{x})}{\partial
t}=-i\vec{\alpha}\cdot\vec{\nabla}\psi(t,\vec{x})+m\beta\psi(t,\vec{x})~.
\end{equation*}
Terms containing $\beta$ cannot contribute (it amounts to take the
real part of an imaginary number). Thus we obtain the following
current for the i-th coordinate
($\vec{J}(t,\vec{X})=(\vec{j}_1(t,\vec{X}),\ldots,\vec{j}_\omega(t,\vec{X})$):
\begin{equation*}
\vec{j}_i(\vec{x}_1,\cdot\cdot,\vec{x}_\omega,t)=\sum_{s_i,a_1,\cdot\cdot,a_\omega}
\Psi^{*}_{a_1\cdot\cdot a_i \cdot\cdot
a_\omega}(t,\vec{x}_1,\cdot\cdot,\vec{x}_\omega)
\vec{\alpha}_{{a_i}{s_i}} \Psi_{a_1\cdot\cdot {s_i} \cdot\cdot
a_\omega}(t,\vec{x}_1,\cdot\cdot,\vec{x}_\omega)~,
\end{equation*}
where
\begin{equation*}
\Psi_{a_1\ldots a_\omega}(t,\vec{x}_1,\ldots,\vec{x}_\omega)=
\langle 0_2|\psi_{a_\omega}(\vec{x}_\omega)\ldots
\psi_{a_1}(\vec{x}_1)|\Psi(t)\rangle~.
\end{equation*}
That particular form of the current is unchanged when there is
an interaction term $g(\bar{\psi}\psi)^2$.
\section{\label{sec:dbm}A deterministic Bell model}
We have already mentioned the problem with the continuum limit: the fermion-number contained in any physical state (any state
having a finite number of electrons and positrons) becomes
infinite. So the fermion-number density is infinite too. In order
to keep everything finite, the ultra-violet cut-off must be
maintained. Nevertheless, the positions of the quanta of the
fermion-number must belong to $\mathbb{R}^3$, otherwise the theory
cannot be deterministic. So, let us see if we can obtain a
deterministic Bell model by pursuing these two lines of thought.

The idea is the following. Take the universe to be a finite cubic box of volume $V$, made of $L$ smaller cubic boxes of volume
$\lambda^3$ (where $\lambda^3$ is very small). The boxes are indexed by a label
\begin{equation*}
l=1,2,\ldots,L~.
\end{equation*}
There can be up to four quanta of the fermion-number in each box. A particular configuration of the fermion-number density is still defined by
\begin{equation*}
n=(n_1,n_2,\ldots,n_L)~.
\end{equation*}
There are four operators that create a quantum of the fermion-number in box $l$; these are denoted by
$\psi^\dagger_a(l)$ ($a\in\{1,2,3,4\}$). And there can be up to $4L$ quanta of the fermion-number in the universe. Again, we take
the pilot-state to be an eigenstate of the fermion-number, according to the super-selection rule, with eigenvalue $\omega$.
So the solution of the Schr\"odinger equation amounts to give the wave-function $\Psi_{a_1\ldots a_\omega}(t,l_1,\ldots,l_\omega)$.

At the level of the beables, the quanta of the fermion-number have well defined positions $\vec{x}_1(t),\ldots,\vec{x}_\omega(t)$ and
we assume that when a quantum of the fermion-number enters the box $l$, it will certainly be detected in box $l$ but its exact
position remains unknown. So we have an $\vec{X}(t)=(\vec{x}_1(t),\ldots,\vec{x}_\omega(t))$ which
determines univocally the configuration $n(t)$ that will be observed, if a measurement of the fermion-number density is
performed at time $t$. The determination is made with the formula
\begin{equation*}
n_l(t)=\sum_{j=1}^{j=\omega}(1\,\,\mathrm{if}\,\,\vec{x}_j(t)\cap\mathrm{box\, l} \neq \emptyset)~.
\end{equation*}
The next step is to extend the domain of definition of the wave-function $\Psi_{a_1\ldots a_\omega}(t,l_1,\ldots,l_\omega)$,
in order to obtain a wave-function $\Psi_{a_1\ldots a_\omega}(t,\vec{x}_1,\ldots,\vec{x}_\omega)$ defined on
$\Rset^{3\omega}$ that can guide the beables $\vec{x}_1(t),\ldots,\vec{x}_\omega(t)$ in the configuration space
$\Rset^{3\omega}$. To do that, we replace the fields $\psi^\dagger_a(l)$ by the operators
\begin{equation*}
\frac{1}{\lambda^3}\displaystyle\int_{\text{box
l}}\psi^\dagger_a(\vec{x})d^3\vec{x}~,
\end{equation*}
where $\psi^\dagger_a(\vec{x})$ is defined at eq. (\ref{sf}). For example, the eigenstate of the fermion-number density
$\psi^\dagger_1(l)|0_2\rangle$ would become
\begin{equation*}
\frac{1}{\lambda^3}\displaystyle\int_{\text{box
l}}d^3\vec{x}\psi^\dagger_1(\vec{x})|0_2\rangle~.
\end{equation*}
Then the complete description of the universe at time $t$ would be given by the couple $(n(t),|\Psi(t)\rangle)$, which is equivalent
to $(\vec{X}(t),|\Psi(t)\rangle)$, since $n(t)$ is fixed by $\vec{X}(t)$. The next step is to give the equations of motion for
these quantities. For the pilot-state, the Schr\"odinger equation is retained. For the vector $\vec{X}(t)$, we must choose a law
that ensures that the predictions of the orthodox quantum field theory are regained. Let us consider a set of universes, labelled
by an index $j$, with the same pilot-state. We can define a probability density for the universe to be in configuration
$\vec{X}$ at time $t$: we call it $r(t,\vec{X})$. In order to regain the predictions of the orthodox theory, it is sufficient to
say that the following condition
\begin{equation}\label{cond}
r(t,\vec{X})=\rho(t,\vec{X})
\end{equation}
must be satisfied ($\rho(t,\vec{X})$ is defined at eq. (\ref{pd})). That ensures that the condition (\ref{equi}), which is
\begin{equation*}
P_n(t)=\sum_{q}|\langle n,q|\Psi(t)\rangle|^2~,
\end{equation*}
is also satisfied. Let us assume that the initial configurations $\vec{X}_j(t_0)$ are chosen according to the probability density
\begin{equation*}
r(t_0,\vec{X})=\rho(t_0,\vec{X})~.
\end{equation*}
Then the condition (\ref{cond}) is equivalent to
\begin{equation}\label{cons}
\frac{\partial r(t,\vec{X})}{\partial
t}=\frac{\partial\rho(t,\vec{X})}{\partial t}~.
\end{equation}
We suppose that the universe moves in a deterministic way; then its velocity $\vec{V}(t)$ must be obtained from the quantities
$|\Psi(t)\rangle$ and $\vec{X}(t)$. We also have the continuity equation
\begin{equation}\label{v1}
\frac{\partial r(t,\vec{X})}{\partial t}+\vec{\nabla}\cdot
(r(t,\vec{X})\vec{V}(t,\vec{X}))=0~.
\end{equation}
With the help of eq.~(\ref{co}) and eq.~(\ref{v1}), eq.~(\ref{cons}) becomes
\begin{equation*}
\vec{\nabla}\cdot(r(t,\vec{X})\vec{V}(t,\vec{X}))=\vec{\nabla}\cdot\vec{J}(t,\vec{X})~.
\end{equation*}
Thus if take the velocity
\begin{equation*}
\vec{V}(t)=\frac{\vec{J}(t,\vec{X})}{\rho(t,\vec{X})}\bigg|_{\vec{X}=\vec{X}(t)}~,
\end{equation*}
all the predictions of the orthodox theory are regained.

The Bell model is non-local, but this is a necessary property of any realistic interpretation of the quantum field theory,
following the EPR paradox, Bell's inequality and related experiments. To show it explicitly, one can consider the case of
two electrons in a $1+1$ space-time. These electrons are described by the beables $x_1(t)$ and $x_2(t)$ and they move according to
the velocity-law
\begin{align*}
&v_1(t)=\frac{j_1(t,x_1,x_2)}{\rho(t,x_1,x_2)}&
&v_2(t)=\frac{j_2(t,x_1,x_2)}{\rho(t,x_1,x_2)}&
\end{align*}
Due to the exchange interaction, required by the Pauli principle, it can be shown that the current $j_1(t,x_1,x_2)$ cannot be
factorized (it is impossible to find two functions $j_A$ and $j_B$ such that $j_1(t,x_1,x_2)=j_A(t,x_1)j_B(t,x_2)$), so the Bell
model is non-local; the velocity of an electron, at time $t$, depends on the position of the other electron, at the same time.
Since the velocity-law has the same expression as that given by Bohm for Dirac particles (first quantization), the velocity-law is
not covariant (see \cite{holland2} for a discussion of the Lorentz covariance of the velocity-law).

There is another deterministic interpretation of the fermionic quantum field theories, due to
Holland (\cite{holland}, section 10.6.2), where the fermionic field is treated as a collection
of rotators. For the simplest case
of a spin $0$ field quantized according to the Fermi-Dirac statistics, there is a rotator for each
normal mode $\vec{k}$ ($\vec{k}=\frac{2\pi}{V^{1/3}}(n_1,n_2,n_3)$,
with $n_1,n_2,n_3\,\in\Rset^3$). For each rotator, there are two independent states,  $u_{\vec{k}-}(\vec{\alpha}_{\vec{k}})$ and
$u_{\vec{k}+}(\vec{\alpha}_{\vec{k}})$, respectively of spin down and spin up, $\vec{\alpha}_{\vec{k}}$ being the Euler
angles of that rotator. If the rotator $\vec{k}$ is in the spin down state, there is no particle of momentum $\vec{k}$ in the universe and conversely if
the rotator is in the spin up state, there is a particle of momentum $\vec{k}$ in the universe.
In Holland's model, the hidden variables are the Euler angles $\vec{\alpha}_{\vec{k}}(t)$, $\vec{\alpha}_{\vec{k}}(t)$ determining univocally if there is a particle of momentum
$\vec{k}$ in the universe at time $t$. Each $\vec{\alpha}_{\vec{k}}(t)$ evolves according to a deterministic law;
$\dot{\vec{\alpha}}_{\vec{k}}(t)$ depends on the other Euler angles (at the same time) and on the state $|\Psi(t)\rangle$.
The extension to a Dirac field seems direct; the number of hidden
variables would be multiplied by four (positron, electron and their two states of spin). How does this model relate to the Bell model?
In Holland's model, which relies on a particle
ontology, the specification of the Euler angles determine the number of particles present in the universe and their velocities. Clearly, Holland's model
has the virtue to show that it is possible to give an objective and
deterministic interpretation of any fermionic quantum field-theoretic model based on a particle ontology, but the price to pay is that the hidden
variables
belong to the momentum space. But it seems to be a necessary price; once we look at localized properties, we have to expect that these properties
will not commute with the particle number. Nevertheless, since any measurement is finally a measurement of positions, the Euler angles are not
revealed. In the Bell model, a measurement of the charge density reveals a pre-existing value, but the link with particles is lost.
Both models rely on a quantum-based ontology (quanta of the particle number in Holland's model, quanta of the fermion number in Bell's model), something which seems justified by the pilot-wave theory of non-relativistic quantum mechanics.

Later, another realistic (and still deterministic) interpretation of the fermionic quantum field theory, due to Valentini \cite{valentini}, has been advanced.
It is the logical completion of Bohm's program for the quantum field theory (fields as beables). Is there some relation between the Bell model
and Valentini's model? In \cite{valentini}, a Van der Waerden field $\phi(t,\vec{x})$ is used, instead of the usual Dirac field. It is
a two-component complex field satisfying the equation
\begin{equation*}
\frac{\partial^2\phi(t,\vec{x})}{\partial t^2}-
(\vec{\sigma}\cdot\vec{\nabla})^2\phi(t,\vec{x})+m^2\phi(t,\vec{x})=0~.
\end{equation*}
The Van der Waerden quantum field theory  is supposed to be equivalent to the Dirac quantum field theory. In the Valentini
model, the fields $\phi_a(t,\vec{x})$ and $\phi^*_a(t,\vec{x})$ ($a=1,2$) are the beables. The point we want to make here is that
this model and the Bell model are not equivalent. For example, in the one-quantum case, a field cannot mimic the beable
$\vec{x}(t)$. Even if the field is localized around $\vec{x}(t_0)$ at the initial time $t_0$, there are solutions of the pilot-state
that will make the field spread. This is shown in \cite{valentini}, when the non-relativistic limit is studied and
when the most probable field configurations are deduced, but also in \cite{colin0}, from the velocity-law.

The Bell model has also been studied in \cite{sudbery}, where it is shown that for a non-relativistic hamiltonian
\begin{equation*}
H_{n-r}=\sum_{s}\int
d^3\vec{x}~C^\dagger_s(\vec{x})[-\frac{\Delta^2}{2m}+V(\vec{x})]C_s(\vec{x})~,
\end{equation*}
where $C_s(\vec{x})=\int d^3\vec{p}~c_s(\vec{p})e^{i\vec{p}\cdot\vec{x}}$, and for the case
where there is only one electron, the Bell model is equivalent to the non-relativistic de Broglie-Bohm model. Since it is assumed
that there are only electrons of positive energy, the fermion-number density commutes with the particle-number, but this
only valid for that non-relativistic model. Intuitively, when there are only low-energy electrons (positive-energy electrons),
it seems indeed reasonable that they cannot excite the electrons of the Dirac sea, so that the positive-energy electrons decouple
from the negative-energy electrons, but a complete study should also take account of the potential energy (virtual particles).
\section{Conclusion}
We have shown that the Bell model is not a model concerned about
the trajectories of the particles; the real beable is the charge
density. And it does not commute with the particle number. That
stems from general physical arguments: to measure localized
properties with high precision, one has to use high energy and
that leads to pairs creation.

Can one build a similar interpretation of the Klein-Gordon theory?
It seems that the answer is no, for it is impossible to define a
state destroyed by a charge annihilator in the Klein-Gordon
theory. But this is not even necessary since all measuring devices
are made of fermions. Other observables could be given the beable
status, provided that they are not forbidden by no
hidden-variables theorems, but it seems that only conserved
quantities must be chosen for the model to remain deterministic.
\begin{ack}
The author would like to thank Professor Jean Bricmont and Doctor
Thomas Durt, for taking the time to discuss some of these ideas
with him. He would like also to thank Professor Tony Sudbery, for
his appreciation of an earlier version of this work \cite{colin}.
\end{ack}

\end{document}